\documentclass{jpp}

\usepackage[utf8]{inputenc}
\usepackage[T1]{fontenc}
\usepackage{amsfonts,amsmath,amssymb}
\usepackage{bm,nicefrac}
\usepackage{graphicx}
\usepackage{tabularx}
\usepackage{subcaption}
\usepackage[section]{placeins}
\usepackage{textgreek}
\usepackage[breaklinks=true,colorlinks=true,linkcolor=blue,urlcolor=blue,citecolor=blue]{hyperref} 
\usepackage[table]{xcolor}
\usepackage[english]{babel}
\usepackage{listings}

\title{High Order Free Boundary MHD Equilibria in DESC}
\author{
Rory Conlin \aff{1}  \corresp{\email{wconlin@princeton.edu}}, 
Jonathan Schilling \aff{2}
Daniel W. Dudt \aff{1}, 
Dario Panici \aff{1},
Rogerio Jorge \aff{3},
\and Egemen Kolemen \aff{1} \corresp{\email{ekolemen@princeton.edu}}}

\affiliation{\aff{1}Princeton University, Princeton, New Jersey 08544,
\aff{2}Proxima Fusion, Munich, Germany
\aff{3}Department of Physics, University of Wisconsin-Madison, Madison, Wisconsin 53706, USA}

\begin{document}

\maketitle

\begin{abstract}
In this work we consider the free boundary inverse equilibrium problem for 3D ideal MHD. We review boundary conditions for both fixed and free boundary solutions and under what circumstances a sheet current may exist at the plasma-vacuum interface. We develop an efficient and accurate algorithm for computing the residual of these boundary conditions and use it to compute free boundary equilibria in the DESC code both in vacuum and at finite plasma beta, with and without sheet currents.
\end{abstract}

\section{Introduction}
Free boundary\footnote{Here and throughout we follow the convention in MHD equilibrium and use "free boundary" to mean the case where the boundary shape of the plasma is allowed to change, though the computational boundary is often fixed. This is in contrast to traditional fluid modelling where free boundary refers to the computational boundary being allowed to move.} 3D MHD equilibrium calculations are a key step in many areas of fusion research. For experiments, such calculations are required to reconstruct the internal behavior of the plasma from external measurements \cite{hanson2009v3fit,schilling2018experimental}, as well as in experimental planning when to determine coil currents. Similarly in design they are an important verification step when designing coils for a new equilibrium. In stellarator optimization, free boundary calculations are generally required when performing "single stage" optimization where both the plasma and coils are optimized at the same time to achieve specific objectives \cite{henneberg2021combined,drevlak_onset_private_comm}.

The primary code presently used for free boundary calculations is VMEC \cite{hirshman_steepestdescent_1983, hirshman_three-dimensional_1986}, which uses the NESTOR code \cite{merkel_integral_1986,merkel_1988,merkel2015linear} to compute the required vacuum field outside the plasma. NESTOR solves the exterior Neumann problem using a Green's function method, and a singularity subtraction scheme to solve the resulting singular integrals. This method is only 2nd order accurate in the number of points used to discretize the integral, consistent with the 2nd order accuracy of VMEC's radial discretization. 

The DESC code \cite{desc_git} improves upon VMEC in a number of ways, and has been shown to be significantly more accurate than VMEC \cite{panici2023desc}, though is currently limited to fixed boundary calculations. In extending DESC to free boundary, it is natural to seek a method that will preserve this improved accuracy. Recent work by Malhotra \cite{malhotra2019taylor, malhotra_efficient_2020} has developed a new "partition of unity" method for computing singular boundary integrals arising in MHD equilibrium calculations that has been shown to be much more accurate, achieving 10th to 12th order accuracy \cite{malhotra_efficient_2020}, and is the method implemented in DESC and discussed in this work. We note that NESTOR has been ported to Python/JAX and interfaced with DESC and is available for comparisons. We begin by reviewing the boundary conditions for a fixed boundary problem in \autoref{sec:fixedbc} and the free boundary problem in \autoref{sec:freebc}. We then move on to the formulation of these boundary conditions that is solved in DESC in \autoref{sec:method} and the details of the numerical implementation in \autoref{sec:numerics}. We finally show benchmarks of free boundary DESC vs both VMEC and direct field line tracing in \autoref{sec:results}.

\section{Fixed Boundary Conditions}
\label{sec:fixedbc}

Before considering the free boundary problem, it is helpful to first consider the fixed boundary problem, in which the shape of the last closed flux surface ($R_b, Z_b$) is specified, along with the normal field (taken to be zero). When a code such as DESC or VMEC is run with a fixed boundary, the magnetic field it finds is the \textit{total} magnetic field in the plasma volume due to all sources, not just the currents in the plasma volume. This total field $\mathbf{B}_\textrm{fixed}$ can be thought of as being made up of two parts as $\mathbf{B}_\textrm{fixed} = \mathbf{B}_\textrm{plasma} + \mathbf{B}_\textrm{fake}$, where $\mathbf{B}_\textrm{plasma}$ is the field due to the volume currents in the plasma and $\mathbf{B}_\textrm{fake}$  is a "fake" external field that is required to make the last closed flux surface coincide with the specified boundary. Because the source of the "fake" field is not inside the plasma volume, within the plasma volume it is curl free: $\nabla \times \mathbf{B}_\textrm{fake} = 0$.

A common physical model for such a configuration is that of a plasma surrounded by a rigid superconducting shell.
Because the magnetic field~$\mathbf{B}_\textrm{fixed}$ inside a ideal superconductor is zero, the boundary conditions for Maxwell's equations in matter give

\begin{equation}
    \mathbf{K}_\textrm{SC} = -\frac{1}{\mu_0} \mathbf{n}  \times \mathbf{B}_\textrm{fixed}  
\end{equation}

where $\mathbf{n}$ is the surface normal vector and
$\mathbf{K}_\textrm{SC}$ is a sheet current density that exists on the plasma/superconductor interface, cancelling the field due to currents in the plasma volume such that the total field outside the plasma is zero, and being the source of the "fake" field inside the plasma.

The goal of coil design in the second stage of a stellarator design process is to find external coils that replicate the effect of this sheet current inside the plasma volume - effectively to replace the "fake" field in the plasma volume with the field from external coils. With the plasma boundary held fixed and the normal field from the plasma $\mathbf{B}_\textrm{plasma} \cdot \mathbf{n}$ specified, the solution inside the plasma volume is unique, and so any set of coils that lie outside the plasma volume and give $(\mathbf{B}_\textrm{plasma} + \mathbf{B}_\textrm{coil}) \cdot \mathbf{n} = 0$ will give the same equilibrium in the plasma volume.

The question may then be asked: what happens to the sheet current? If one performs a Biot-Savart integral over the plasma currents in the volume to find $\mathbf{B}_\textrm{plasma} \cdot \mathbf{n}$, the effect of the sheet current will not be included, and so properly designed coils will create the original equilibrium with no sheet current present. The same is generally true when the volume integral is transformed into a surface integral using the virtual casing principle - in fact we see that the virtual casing sheet current $\mathbf{K}_\textrm{VC} = -\frac{1}{\mu_0} \mathbf{n}  \times \mathbf{B}_\textrm{fixed}$ is exactly the physical sheet current that exists in a fixed boundary equilibrium. In general, if we first obtain a fixed boundary equilibrium and then design coils for it, the free boundary equilibrium from the given coils will not have any sheet current on the plasma-vacuum interface.

\section{Free Boundary Conditions}
\label{sec:freebc}

The boundary conditions for a plasma supported by external fields can be found in many standard texts on MHD \cite{freidberg2014ideal, goedbloed2019magnetohydrodynamics}, and are summarized below

\begin{equation}
    \mathbf{B} \cdot \mathbf{n} = 0
    \label{eq:bn}
\end{equation}

\begin{equation}
    [[p + B^2/2 \mu_0 ]] = 0
    \label{eq:b2}
\end{equation}

\begin{equation}
    \mathbf{n} \times [[\mathbf{B}]] = \mu_0 \mathbf{K}
    \label{eq:nbk}
\end{equation}

where $\mathbf{B}$ is the magnetic field, $p$ is the plasma pressure, $\mathbf{K}$ is a surface current density, $\mathbf{n}$ is the normal vector to the plasma boundary, and $[[x]]$ is the jump in $x$ across the boundary. Because $\mathbf{B}$ lies tangent to the surface, $\mathbf{K}$ will as well. These conditions can be derived intuitively by noting that if the first were not satisfied, particles streaming along magnetic field lines could leave the domain, while if the second were not satisfied the pressure difference would cause the plasma boundary to move, in both cases the plasma would not be in equilibrium. The third equation is simply a consequence of Maxwell's equations for a conductor such as a plasma immersed in an external magnetic field.

From \autoref{eq:b2}, we see that if the plasma pressure at the edge is nonzero, then there must be a corresponding jump in the magnetic pressure $B^2/2\mu_0$. The first boundary condition requires that the normal field is continuous (and zero) across the interface, so there must be a jump in the tangential field. This then requires the existence of a sheet current on the plasma-vacuum interface. In the more common case that the plasma pressure does go to zero at the edge, then there is no explicit need for a sheet current, but neither is it forbidden. Although both the normal field and magnetic pressure are continuous in such a case, a sheet current can still exist that rotates the tangential field without affecting its magnitude.

To determine if a sheet current exists in such a case, we can consider an equilibrium located such that the combined field~$\mathbf{B}$ from the plasma and the external coils with $\mathbf{B} = \mathbf{B}_\textrm{plasma}  + \mathbf{B}_\textrm{coil}$ has zero normal component, so \autoref{eq:bn} is satisfied. Because both $\mathbf{B}_\textrm{plasma}$ and $\mathbf{B}_\textrm{coil}$ are continuous everywhere, there can be no jump in the magnetic field across the boundary, so \autoref{eq:b2} is also satisfied, and hence \autoref{eq:nbk} is satisfied with $\mathbf{K} = 0$. This is then the unique solution to the problem. The fact that the normal field error acts as a source term for the sheet current was previously derived more rigorously by Hanson \cite{hanson2016surface, hanson2017surface}. As discussed above, for an equilibrium supported by well designed matching coils, the normal field error and hence the sheet current will be negligible. However, there are still many common cases where a sheet current will exist, such as a finite beta equilibrium supported by coils designed for a vacuum equilibrium or vice versa.

\section{Method of Solution}
\label{sec:method}

We now consider the incorporation of the free boundary conditions into a fixed boundary inverse equilibrium solver such as DESC. Inside the plasma, we have $\mathbf{B} = \mathbf{B}_\textrm{fixed}$ where $\mathbf{B}_\textrm{fixed}$ is the magnetic field from a fixed boundary equilibrium solution, which inherently satisfies $\mathbf{B}_\textrm{fixed} \cdot \mathbf{n} = 0$ on the surface. This field is made of 2 components:

\begin{equation}
  \mathbf{B}_\textrm{fixed} = \mathbf{B}_\textrm{plasma} + \mathbf{B}_\textrm{fake}  
\end{equation}

where $\mathbf{B}_\textrm{plasma}$ is the field due to the volume currents in the plasma and $\mathbf{B}_\textrm{fake}$ is a "fake" external field that makes $\mathbf{B}_\textrm{fixed} \cdot \mathbf{n} = 0$ on the surface.

In reality, the total magnetic field~$\mathbf{B}$ everywhere is due to 3 different sources: External coils, plasma, and surface currents at the plasma vacuum interface:

\begin{equation}
    \mathbf{B} = \mathbf{B}_\textrm{coil} + \mathbf{B}_\textrm{plasma} + \mathbf{B}_{K}    
\end{equation}

$\mathbf{B}_\textrm{coil}$ is assumed known everywhere. To find $\mathbf{B}_\textrm{plasma}$ we can do a Biot-Savart integral over the volume current density $\mathbf{J}$, or a virtual casing integral over the surface of the virtual casing shielding current $\mathbf{n} \times \mathbf{B}_\textrm{fixed}$. The physical surface current $\mathbf{K}$ is given as $\mathbf{K} = \mathbf{n} \times \nabla \Phi$ for an unknown current potential $\Phi$ which is to be found along with the boundary shape.

Denote the total field outside the plasma as $\mathbf{B}_\textrm{out}$. We evaluate $\mathbf{B}_\textrm{out}$ by summing the coil field and the fields due to the virtual casing sheet current and the physical sheet current, obtained via a 2D Biot-Savart integral over the boundary:

\begin{equation}
  \mathbf{B}_\textrm{out} = \mathbf{B}_\textrm{coil} + \frac{\mu_0}{4\pi}\int_D \frac{\mathbf{n} \times (\mathbf{B}_\textrm{fixed} + \nabla \Phi) \times (\mathbf{r} - \mathbf{r}')}{|\mathbf{r} - \mathbf{r}'|^3} \mathrm{d}^2\mathbf{r}'  
  \label{eq:bout}
\end{equation}

With the inner~($\mathbf{B}_\textrm{fixed}$) and outer~($\mathbf{B}_\textrm{out}$) fields thus specified, we can then define the problem:

\begin{equation}
    \min_{R_b, Z_b, \Phi} 
\begin{pmatrix}
    \mathbf{B}_\textrm{out} \cdot \mathbf{n} \\
    \frac{1}{2 \mu_0} B^2_\textrm{out} - \frac{1}{2 \mu_0} B^2_\textrm{fixed} - p \\
    \mathbf{n} \times \nabla \Phi - \mathbf{n} \times (\mathbf{B}_\textrm{out} - \mathbf{B}_\textrm{fixed}) \\
    \label{eq:objective}
\end{pmatrix}
\end{equation}

where $\mathbf{B}_\textrm{out}$ and $\mathbf{B}_\textrm{fixed}$ are implicitly functions of the boundary shape $R_b, Z_b$. We have also taken advantage of the fact that $p$ outside the boundary is zero, and $\mathbf{B}_\textrm{fixed} \cdot \mathbf{n}$ is identically zero by construction of the fixed boundary field. In DESC, these equations are implemented as an objective function that should be minimized by varying the boundary shape, subject to the constraint that the plasma is in force balance. In practice the equations are solved using a constrained quasi-Newton method described in \cite{conlin2023desc, dudt2023desc}.

\section{Numerical Technique}
\label{sec:numerics}

The fixed boundary equilibrium is obtained from running DESC in fixed boundary mode for a particular guess of the boundary shape, which furnishes $\mathbf{B}_\textrm{fixed}$. The numerical details of DESC can be found in a series of papers \cite{dudt2020desc, panici2023desc, conlin2023desc, dudt2020desc}. 

The external field from the coils can be specified in a number of ways. For backwards compatibility with VMEC, we implement an interface to MGRID files where the field from the coils is splined over a grid in lab coordinates $(R, \phi, Z)$ and interpolated to the current boundary position. This is a relatively fast method of evaluating the external field, but is only first order accurate in the grid spacing. Another method is to directly compute the field from a set of filamentary coils using the Biot-Savart law, which is the most accurate though somewhat slower. Both of these methods are used for the results of \autoref{sec:results}. In general, the code is flexible enough such that any Python function capable of evaluating the external field at a point in lab coordinates can be provided by the user.

With the fixed boundary field and the coil field specified, the only remaining difficulty is in evaluating the singular Biot-Savart integral in \autoref{eq:bout} over the surface. To do so, we adapt the high order singular integration technique of Malhotra et al \cite{malhotra_efficient_2020, malhotra2019taylor} which we summarize below.

We consider a toroidal surface parameterized by poloidal angle $\theta$ and toroidal angle $\zeta$, with metric tensor determinant $g$ and we seek to evaluate the integral over the surface of a singular kernel $h$

\begin{equation}
    F(\theta, \zeta) = \int_0^{2\pi} \int_0^{2\pi} h(\theta, \zeta, \theta', \zeta') g(\theta', \zeta') d\theta' d\zeta'
\end{equation}

We further assume that the values of $h$, $g$, $\mathbf{r}$ are known on a regular grid in $\theta'$ and $\zeta'$ with nodes $\theta'_i$, $\zeta'_i$ and spacing $\Delta\theta'$, $\Delta\zeta'$. 

We first split the integrand into two parts, a globally defined smooth part and a singular part with finite support:

\begin{align}
    F(\theta, \zeta) =&\, \phantom{+}\, \int_0^{2\pi} \int_0^{2\pi} h_\textrm{smooth}(\theta, \zeta, \theta', \zeta') g(\theta', \zeta') d\theta' d\zeta' \nonumber \\
                     ~&\,          +    \int_0^{2\pi} \int_0^{2\pi} h_\textrm{singular}(\theta, \zeta, \theta', \zeta') g(\theta', \zeta') d\theta' d\zeta'
\end{align}

where 

\begin{equation}
    h_\textrm{smooth} = h \, (1 - \chi(\rho))
\end{equation}

and 

\begin{equation}
    h_\textrm{singular} = h \, \chi(\rho)
\end{equation}

$\rho$ is the normalized distance between the source and evaluation points:

\begin{equation}
    \rho = \frac{2}{s} \sqrt{\Bigg(\frac{\theta - \theta_i'}{\Delta\theta}\Bigg)^2 + \Bigg(\frac{\zeta - \zeta_i'}{\Delta\zeta}\Bigg)^2}
\end{equation}
with $s$ determining the size of finite support of the singular domain. $\chi$ is a "partition of unity" that splits the domain into the smooth and singular parts. We follow Malhotra and choose 

\begin{equation}
    \chi(\rho) = e^{-36 \rho^8}
\end{equation}

though formally any function that goes to zero for $\rho > 1$ and is equal to 1 in a neighborhood of zero would work.

Because $\chi = 1$ in a neighborhood of the origin, it fully suppresses the singularity in $h$ such that $h_{smooth}$ is a smooth function on a periodic domain for which standard trapezoidal integration will converge exponentially.

To integrate the singular part, we transform to polar coordinates $(\rho, \omega)$ around the singular point. The Jacobian of this transformation is $\rho$ which cancels singularities of the $1/r$ type, and it was shown \cite{ying2006boundary_integral} that the scheme yields the correct principal value for hypersingular kernels such as the Biot-Savart kernel, provided the quadrature rule used is symmetric about the origin. This allows the use of standard trapezoidal quadrature in $\omega \in (0, \pi)$ and Gauss-Legendre quadrature in $\rho \in (-1, 1)$.

To evaluate the integral in polar coordinates, we must interpolate values from the regular grid in $(\theta, \zeta)$ to the polar coordinates around each singular point $(\rho_i, \omega_i)$. In the original work Malhotra et. al. used 12th order Lagrange polynomial interpolation from a 12x12 rectangular grid surrounding each polar coordinate, though they note that with this method interpolation of derived quantities such as the metric tensor were inaccurate, and so the basis vectors themselves were interpolated and the metric tensor recalculated in polar coordinates. We instead take advantage of the fact that in the most common case where we need to evaluate the integral on a regular grid in $(\theta, \zeta)$, then the polar points surrounding each evaluation point are also in a regular grid, albeit shifted with respect to the base grid. This allows us to use Fourier interpolation, accounting for the shift by multiplying in the Fourier domain by a complex phase factor. This means that the interpolation is spectrally accurate, and so we find we do not need to recompute any quantities in polar coordinates.

For comparison, VMEC uses the NESTOR code to solve the exterior Neumann problem to find $\mathbf{B}_\textrm{vac} = \mathbf{B}_\textrm{coil} + \mathbf{B}_K$ such that $\mathbf{B}_\textrm{vac}\cdot \mathbf{n} = 0$. This automatically satisfies \autoref{eq:bn} and implicitly satisfies \autoref{eq:nbk} (though the sheet current is never explicitly calculated), but must be iterated to solve \autoref{eq:b2}. The drawback to such an approach is that solving the exterior Neumann problem on the boundary requires inverting a large linear operator whose elements involve multiple singular integrals over the boundary surface, which are expensive to calculate. Our method requires only one evaluation of a singular integral (the virtual casing integral), and avoids any inversion of large linear systems. It is possible to avoid singular integrals entirely by solving the exterior Neumann problem in a volume outside the plasma boundary rather than on the surface, as is done in the SPEC code \cite{hudson_free-boundary_2020}, however this greatly increases the size of the linear system that must be solved, and requires an extra computational boundary outside the plasma but inside the coils to be specified. In some cases such a computational boundary may be easy to determine, but in the case of single stage stellarator optimization where both the plasma boundary and coil geometry is changing, finding a computational boundary that lies in between the two is non-trivial, so we elect to pay the price of the singular integral to alleviate this problem.

\section{Results}

\begin{figure}
    \centering
    \includegraphics[width=1\linewidth]{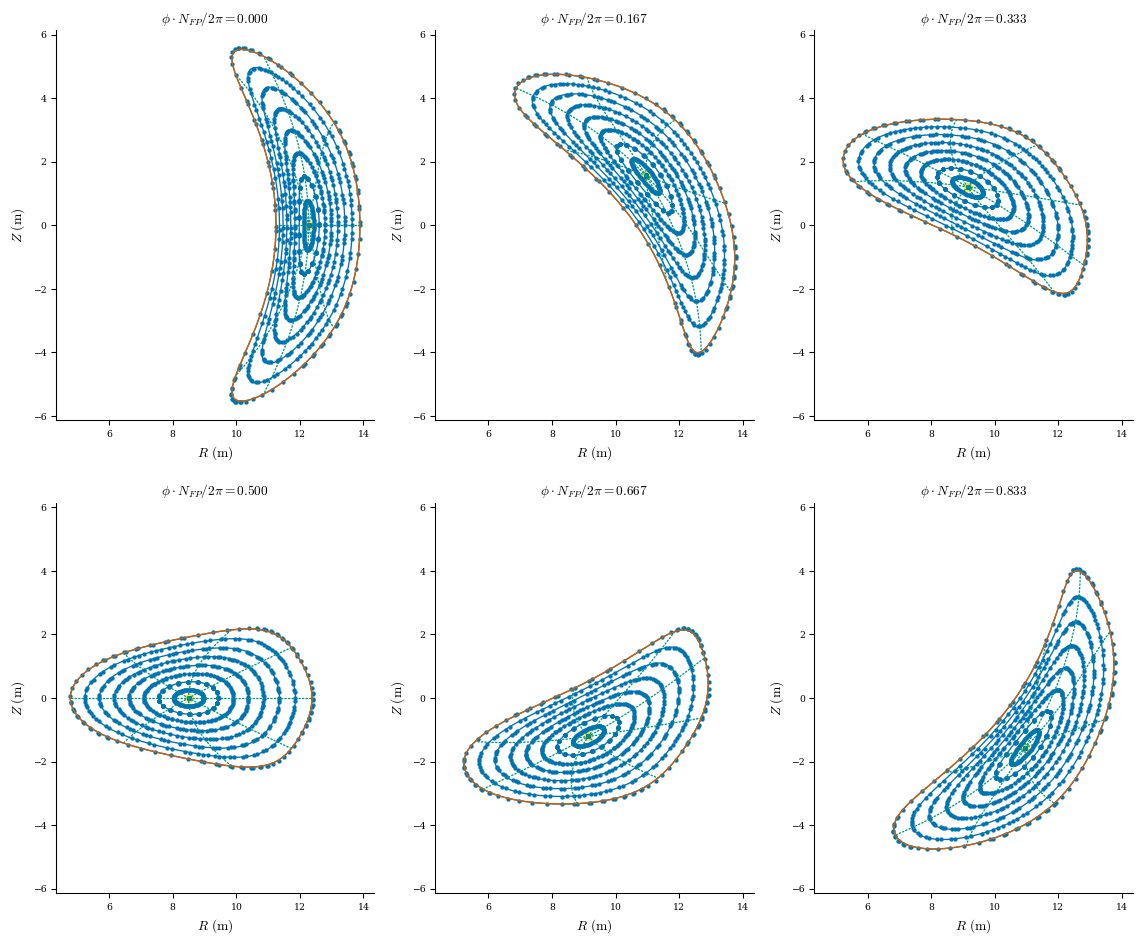}
    \caption{Flux surfaces for the Landreman \& Paul QA equilibrium from DESC, overlayed with field line trajectories traced from the coil set.}
    \label{fig:lpqa_poincare}
\end{figure}

As a first demonstration we use the Landreman \& Paul precise QA equilibrium \cite{landreman2022precise}, scaled to the major radius and field strength of AIRIES-CS, and then create filamentary coils for this equilibrium using SIMSOPT \cite{simsopt2021}. We then run DESC in free boundary mode, where the external field is supplied by directly computing the Biot-Savart integral over the coils at each iteration, and using an objective that minimizes the residual of \autoref{eq:objective} (we drop the last equation as we know a sheet current should not exist for a vacuum field). For an initial guess we take the fixed boundary equilibrium and truncate the Fourier spectrum of the boundary representation to $M=1, N=1$, though the results do not seem to depend on the choice of initialization provided it has approximately the correct major and minor radii and does not intersect the coils. To verify the solution, we trace field lines directly from the coil set and compare the flux surfaces of the equilibrium to the Poincare plot of these field lines. As shown in \autoref{fig:lpqa_poincare}, the field line tracing agrees nearly perfectly with the flux surfaces from DESC.

\label{sec:results}

\begin{figure}
    \centering
    \includegraphics[width=1\linewidth]{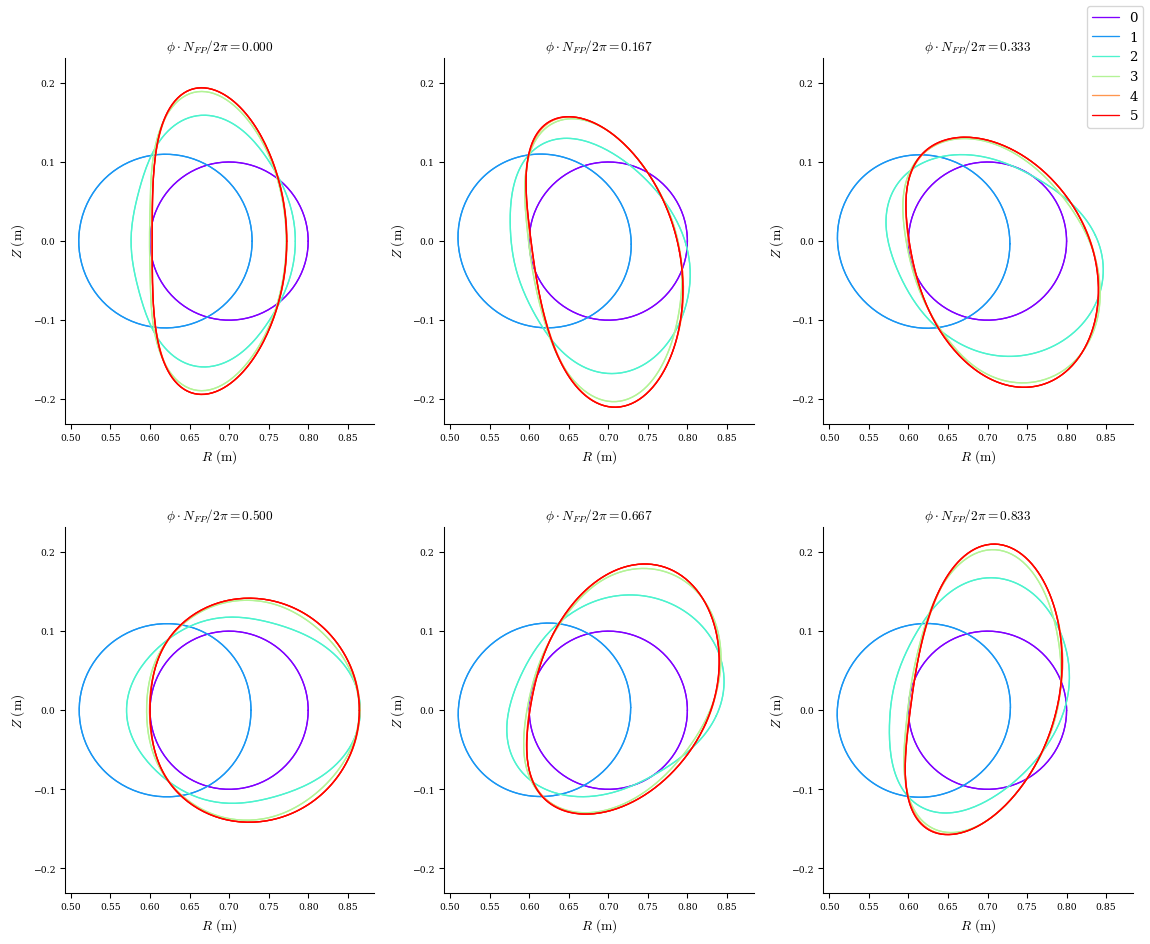}
    \caption{Deformation of initially circular boundary shape into helical stellarator.}
    \label{fig:iterations}
\end{figure}

To show the progression of the solution, we show another example inspired by one shown in the origin free boundary VMEC paper \cite{hirshman_three-dimensional_1986}, where we solve for a helical stellarator starting from an initial guess that is a circular torus with zero current and zero pressure. In this case the external field was provided by MGRID on a grid in lab coordinates and interpolated to the required evaluation points. As shown in \autoref{fig:iterations}, the boundary shape converges to the correct shape after only a handful of iterations, with the solution not changing perceptibly after 4 steps.

\begin{figure}
    \centering
    \includegraphics[width=0.5\linewidth]{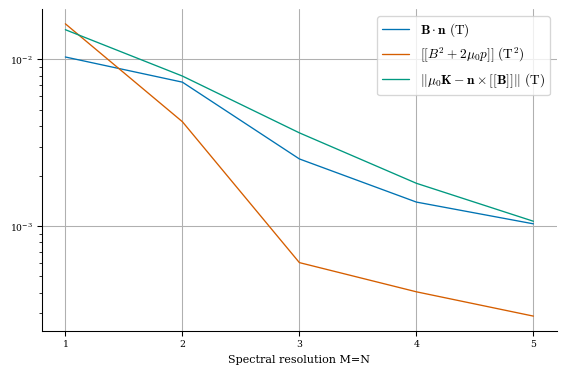}
    \caption{Convergence of boundary condition residual as function of spectral resolution for helical stellarator with finite edge pressure.}
    \label{fig:surf_current_convergence}
\end{figure}

To verify the sheet current calculation, we solve a helical stellarator with $\beta \sim 1\%$ and finite edge pressure which we choose to be half that of the core pressure. In order to satisfy \autoref{eq:b2}, the magnetic field must be discontinuous, leading to a sheet current on the plasma boundary. \autoref{fig:surf_current_convergence} shows the convergence of the three boundary conditions as the spectral resolution of both the equilibrium and the surface current is increased. We find that the poloidal average of the surface current converges to $\sim 3.6~kA$, approximately $1\%$ of the total poloidal current provided by the coils and plasma.

\begin{figure}
    \centering
    \includegraphics[width=1\linewidth]{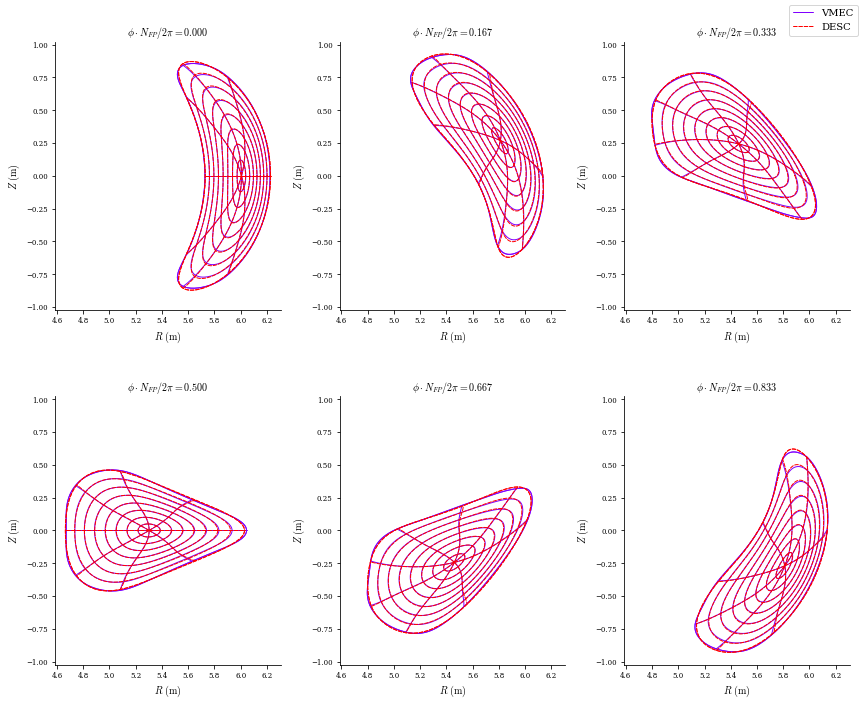}
    \caption{Comparison of free boundary DESC and VMEC solutions for W7-X at $\beta=2\%$.}
    \label{fig:w7x}
\end{figure}

As a last example, we show a comparison to free boundary VMEC for a W7-X like equilibrium at $\beta=2\%$. As shown in \autoref{fig:w7x} we get very good agreement in the flux surfaces near the core, but there are some small differences in the boundary shape, on the order of 4mm, which does not seem to decrease appreciably as resolution is increased. For verification, we computed the residual in \autoref{eq:objective} for both the VMEC and DESC boundaries using the high order method detailed in \autoref{sec:numerics} and find that the DESC boundary has 2-3x lower error. We attribute this to the higher order boundary integral method used in DESC, and is consistent with previous results comparing schemes for singular boundary integrals \cite{malhotra_efficient_2020}.

\section{Conclusion}

In this work we have developed and implemented the capability to compute free boundary equilibria in the DESC code using a recently developed high order method for singular boundary integrals, and avoiding having to explicitly solve the exterior Neumann problem. We have benchmarked this new capability for both vacuum and finite beta equilibria against both field line tracing and free boundary VMEC, and shown that the high order method in DESC appears to be more accurate than VMEC. By casting the boundary conditions as an objective in an optimization problem, extension to single stage optimization where both the plasma and external field are allowed to change is straightforward, and will be the subject of a future publication.

\section*{Acknowledgements}
Thanks to Joachim Geiger for providing the external field for the W7-X benchmark, and to Dhairya Malhotra and Sophia Henneberg for helpful discussions regarding methods for singular integrals and boundary conditions for MHD equilibria.

\section*{Funding}
This work was supported by the U.S. Department of Energy under contract numbers DE-AC02-09CH11466, DE- SC0022005 and Field Work Proposal No. 1019. The United States Government retains a non-exclusive, paid-up, irrevocable, world-wide license to publish or reproduce the published form of this manuscript, or allow others to do so, for United States Government purposes.

\section*{Declaration of interests}
The authors report no conflict of interest.

\section*{Data availability statement}
The source code to generate the results and plots in this study are openly available in DESC at \url{https://github.com/PlasmaControl/DESC} or \url{http://doi.org/10.5281/zenodo.4876504}.

\section*{Author ORCID}
R. Conlin, \url{https://orcid.org/0000-0001-8366-2111}; D. Dudt, \url{https://orcid.org/0000-0002-4557-3529}; D. Panici, \url{https://orcid.org/0000-0003-0736-4360}; E. Kolemen, \url{https://orcid.org/0000-0003-4212-3247}; J. Schilling \url{https://orcid.org/0000-0002-6363-6554}

\bibliographystyle{jpp}

\bibliography{bibliography}

\end{document}